\documentclass[conference,letterpaper]{IEEEtran} 

\usepackage[utf8]{inputenc}
\usepackage[T1]{fontenc}
\usepackage[hidelinks]{hyperref}
\usepackage{url}
\usepackage{ifthen}
\usepackage{cite}
\usepackage[cmex10]{amsmath} 
\usepackage{amsfonts,amssymb}
\usepackage{color}
\usepackage{ifthen}
\usepackage[arxiv]{optional} 

\def\tw{\tilde{w}}
\newcommand{\eq}[1]{\begin{align}#1\end{align}}
\newcommand{\seq}[1]{\begin{subequations}#1\end{subequations}}

\begin{document}
\title{A dynamic program to achieve capacity of multiple access channel with noiseless feedback}

 \author{%
  \IEEEauthorblockN{Deepanshu Vasal}
  \IEEEauthorblockA{Northwestern University\\
                     Evanston, IL 60201, USA\\
                     \texttt{dvasal@umich.edu}}
 }

\maketitle

\begin{abstract}
In this paper, we consider the problem of evaluating capacity expression of a multiple access channel (MAC) with noiseless feedback. So far, the capacity expression for this channel is known through a multi letter directed information by Kramer~\cite{Kr98}. Recently, it was shown in~\cite{AnPr20} that one can pose it as a dynamic optimization problem, however, no dynamic program was provided as the authors claimed there is no notion of state that is observed by both the senders. In this paper, we build upon~\cite{AnPr20} to show that there indeed exists a state and therefore a dynamic program (DP) that decomposes this dynamic optimization problem, and equivalently a Bellman fixed-point equation to evaluate capacity of this channel. We do so by defining a common belief on private messages and private beliefs of the two senders, and using this common belief as state of the system. We further show that this DP can be further reduced to a DP with state as the common belief on just the messages.
This provides a single letter characterization of the capacity of this channel.
\end{abstract}



\def\cE{\mathcal{E}}
\def\cX{\mathcal{X}}
\def\cY{\mathcal{Y}}
\def\cZ{\mathcal{Z}}
\def\cW{\mathcal{W}}
\def\cP{\mathcal{P}}
\def\cU{\mathcal{U}}
\def\cV{\mathcal{V}}
\def\cR{\mathcal{R}}
\def\cC{\mathcal{C}}
\def\cS{\mathcal{S}}
\def\cF{\mathcal{F}}
\def\cG{\mathcal{G}}
\def\cB{\mathcal{B}}

\def\tw{\tilde{w}}

\newtheorem{lemma}{Lemma}
\newtheorem{fact}{Fact}
\newtheorem{theorem}{Theorem}

\newcommand{\ve}[1]{\underline{#1}}
\newcommand{\eqdef}{\stackrel{\scriptscriptstyle \triangle}{=}}
\newcommand{\mdef}{\stackrel{\text{\tiny def}}{=}}
\newcommand{\E}{\mathbb{E}}
\def\Real{\mathbb{R}}
\def\P{\mathbb{P}}

\vspace*{-0.15cm}
\section{Introduction}
Finding the capacity and capacity achieving scheme of Multiple Access Channel with noiseless feedback (MAC-NF) is a fundamental problem in communication that has been studied from the lens of information theory~\cite{GaKi11}. It was shown by Gaarder and Wolf~\cite{GaWo75} that feedback strictly increases the capacity of this channel. Later Cover and Leung~\cite{CoLe81} provided an achievable rate region for this channel using block Markov coding and showed it to be tight for a class of channels for which each encoder can perfectly decode the transmitted symbol of the other using feedback and its own transmitted symbol (henceforth referred to as Cover-Leung Channels). Kramer~\cite{Kr98} provided a multi-letter directed information expression for the capacity of this channel. For Gaussian MAC with feedback, Ozarow~\cite{Oz84} provided a linear scheme that achieves capacity and shows that for this channel, the expression for capacity is single letter. Since then finding a single letter expression for MAC channel with feedback has been an open problem.

Recently, Anastasopoulos and Pradhan in~\cite{AnPr20} presented a connection between the problem of Decentralized sequential active hypothesis testing (DSAHT) and that of finding capacity of a multiple access channel with noiseless feedback. DSAHT problem consists of minimizing the terminal probability of error for MAC-NF. For this, the authors show that this can be posed as a decentralized stochastic control problem, and using common information approach~\cite{NaMaTe13}, they pose it as a dynamic program for common agent and show that there exists Markovian strategies that are optimal, where these policies are Markovian with respect to a common belief state $\pi_t$ that puts a belief on private messages of the users. For the problem of achieving capacity, they show that the multi letter expression of capacity defined by Kramer~\cite{Kr98} is a dynamic optimization problem. However, the instantaneous costs involve private beliefs of the senders on their own transmitted symbols, that are not observed by each other. Thus, the instantaneous costs can not be written as a function of a state that is observed by all the players, in effect showing that there is no notion of state and this problem can not decomposed sequentially through as a dynamic program. However, they argue that one can still restrict oneself to Markovian policies that are optimal for DSAHT problem to solve for the dynamic optimization problem which in turn would solve for finding capacity. In summary, their main result shows that there exist a common belief based Markovian strategy that achieves capacity for this channel, and thus one can restrict oneself to this class of strategies to evaluate capacity. This is given by solving a dynamic optimization problem that can not be sequentially decomposed (i.e. does not have a dynamic program).

In this paper, we build upon their work to show that there indeed exists a state and therefore a dynamic program that decomposes this optimization problem, and thus there exists a Bellman fixed-point equation to evaluate capacity of this channel. We do this by again using the common agent approach~\cite{NaMaTe13} and constructing a new common belief state $\tilde{\pi}_t$ that puts a measure on the private messages of the users \emph{as well as} their private beliefs, that appear in the instantaneous cost to achieve capacity, as described above. Based on this new common belief state, we show that this is indeed a state of the system such that it is a controlled Markov process for the problem where the instantaneous cost can written as a function of this state and user's partial functions that map their private state to their transmitted symbols. This shows that one can in principle solve for capacity expression, and the capacity achieving policy, by solving a parameterized Bellman type fixed-point dynamic programming (DP) equation with state as $\tilde{\pi}_t$. We further show that this DP equation can be further simplified to a DP with state as common belief on just the messages of the transmitters i.e. $\pi_t$, the same state considered in~\cite{AnPr20} for the DSAHT problem.
This provides a single letter characterization of the capacity of this channel. 

The paper is structured as follows. Section~\ref{sec:model} provides the model. In Section~\ref{sec:PR}, we discuss previous results. In Section~\ref{sec:DP} we present our main result. In Section~\ref{sec:Conl}, we present conclusion and future work.

\vspace*{-0.15cm}
\section{Channel Model}
\label{sec:model}

We consider a two-user discrete memoryless Multiple Access Channel with Noiseless Feedback (MAC-NF). The input symbols $X^1$, $X^2$ and the output symbol $Y$ take values in the finite alphabets $\cX^1$, $\cX^2$ and $\cY$, respectively.\footnote{We use discrete alphabets to simplify exposition and avoid measure theoretic technicalities that could arise from the continuous alphabets. However, our results go through for continuous alphabets under appropriate technical regularity assumptions.} The channel is memoryless in the sense that the current channel output is independent of all the past channel inputs and the channel outputs, i.e.,
\begin{equation}
\P(y_t|x^1_{1:t}, x^2_{1:t}, y_{1:t-1}) = Q(y_t|x^1_t, x^2_t).
\end{equation}
Our model considers noiseless feedback with unit delay, that is, at time $t$, the presence of the channel outputs $y_{1:t-1}$ to both encoders.

Consider the problem of transmission of messages $m^i\in \mathcal{M}^i=\{1,\ldots,\mathbb{M}^i\}, \; i=1,2$, over the MAC-NF using fixed length codes of length $n$.
Encoders generate their channel inputs based on their private messages and past outputs. Thus
\begin{align}
X^i_t &= \tilde{f}_t^i(M^i,X^i_{1:t-1},Y_{1:t-1})=f_t^i(M^i,Y_{1:t-1}), \quad i=1,2.
\end{align}
The decoder estimates the messages $M^1$ and $M^2$ based on $n$ channel outputs, $Y_{1:n}$ as
\begin{equation}
(\hat{M}^1,\hat{M}^2) = g(Y_{1:n}).
\end{equation}

A fixed-length transmission scheme for the channel $Q$ is the pair $s=(f,g)$, consisting of the encoding functions
$f=(f^1,f^2)$ with $f^i=f^i_{1:n}$ and decoding function $g$.
The error probability associated with the transmission scheme $s$ is defined as
\begin{equation}
P_e(s) = \P^s((M^1,M^2)\neq (\hat{M}^1,\hat{M}^2)).
\end{equation}









\section{DSAHT and MAC-NF channel capacity}
\label{sec:PR}
\subsection{DSAHT}

The problem of Dynamic Sequential Active Hypothesis Testing (DSAHT) involves minimizing the terminal probability of error $P_e(s)$ over all possible transmission schemes $s$, as follows.
Given the alphabets $\mathcal{M}^1,\mathcal{M}^2,\cX$, $\cY$, $\cZ$, the channels $Q^f$, and for a fixed length $T$, design the optimal transmission scheme $s=(f,g)$ that minimizes the error probability $P_e(s)$.
\begin{equation}
P_e = \min_s \P_e(s) \tag{\textbf{P1}}
\end{equation}

For any pair of encoding and update functions, the optimal decoder is the ML decoder (assuming equally likely hypotheses), denoted by $g_{ML}$.

\subsection{Multi-letter capacity expressions}
Shannon derived the capacity for a point to point discrete memoryless channel as maximum of the information between the transmitted symbol and the received symbol~\cite{Sh48}. This is the supremum of the rate below which one can reliably transmit data. However, a similar \emph{single letter} characterization of capacity of MAC-NF is not known.
Kramer in~~\cite{Kr98} provided a multi-letter capacity expression for discrete memoryless MAC-NF, which can be stated as follows.

\begin{fact}[Theorem 5.1 in \cite{Kr98}]
\label{capexp}
The capacity region of the discrete memoryless MAC-NF is
$\cC_{FB}  =\bigcup_{n=1}^{\infty}\cC_{n}$
where $\cC_{n}$, the {directed information $n$-th inner bound region}, is defined as $\cC_n = \text{co}\left(\cR_n\right)$, where $co(A)$ denotes the convex hull of a set $A$, and
\vspace*{-0.25cm}
\begin{align}
\cR_n = \cup_{\cP_n}\{ (R_1,&R_2):
 0 \leq R_1 \leq I_{n}(X^1\rightarrow Z||X^2),  \nonumber \\
& 0 \leq R_2 \leq I_{n}(X^2\rightarrow Z||X^1),\nonumber \\
& 0 \leq R_1+R_2 \leq I_{n}(X^1,X^2\rightarrow Z)
\},
\end{align}
where $I_n(A\rightarrow B||C) = \frac{1}{n}\sum_{t=1}^n I(A_{1:t};B_t|C_{1:t},B_{1:t-1})= \frac{1}{n}\sum_{t=1}^n I(A_{t};B_t|C_{1:t},B_{1:t-1})$.
The above information quantities are evaluated using the joint distribution
\begin{align}
\P(x^1_{1:n},x^2_{1:n},y_{1:n}) = \prod_{t=1}^n Q(y_t|&x^1_t,x^2_t)q^1_t(x^1_t|x^1_{1:t-1},y_{1:t-1})\times \nonumber \\
& q^2_t(x^2_t|x^2_{1:t-1},y_{1:t-1}),
\end{align}
and the union is over all input joint distributions on $x^1_t, x^2_t$ that are conditionally factorizable as
\begin{align}
\label{eq:optdist}
&\P(x^1_t,x^2_t|x^1_{1:t-1},x^2_{1:t-1},y_{1:t-1}) = \nonumber \\
 & \quad q^1_t(x^1_t|x^1_{1:t-1},y_{1:t-1})
 q^2_t(x^2_t|x^2_{1:t-1},Y_{1:t-1})
\end{align}
for $t = 1,2,...,n$.

Furthermore, the regions $\cC_{n}$ can be expressed in the form~\cite{Sa78}
\begin{align}
\label{eq:Tinnerbound}
\cC_n &= \left\{(R_1,R_2)\geq0:\forall\ \underline{\lambda}=(\lambda_1,\lambda_2,\lambda_3)\in \Real^3_+, \right.  \nonumber \\
& \qquad \left. \lambda_1R_1+\lambda_2R_2+\lambda_3(R_1+R_2)\leq C_n(\underline{\lambda})\right\},
\end{align}
where
\begin{subequations}
\label{eq:region_to_cost_fb}
\begin{align}
&C_n(\underline{\lambda}) \triangleq \sup_{\cP_n} I_n(\underline{\lambda})\\
&I_n(\underline{\lambda}) \triangleq
\lambda_1 I_{n}(X^1\rightarrow Z||X^2)+
\lambda_2 I_{n}(X^2\rightarrow Z||X^1)+ \nonumber \\
&\qquad \lambda_3 I_{n}(X^1,X^2\rightarrow Z) \\
 & =\frac{1}{n} \sum_{t=1}^n
 [  \lambda_1 I(X^1_t;Y_t|X^2_{1:t},Y_{1:t-1}) + \nonumber \\
 & \qquad\qquad \lambda_2 I(X^2_t;Y_t|X^1_{1:t},Y_{1:t-1}) +  \lambda_3 I(X^1_t,X^2_t;Y_t|Y_{1:t-1})]
\end{align}
\end{subequations}
and in the above, the set $\cP_n$ is defined as
\begin{align}
\cP_n = \left\{ (q^1_t,q^2_t)_{t=1,\ldots,n}: q^i_t \in  (\cX^i)^{t-1} \times \cZ^{t-1} \rightarrow \cP(\cX^i)   \right\}.
\end{align}
\end{fact}


\subsection{Previous results}
Anastasopoulos and Pradhan in~\cite{AnPr20} considered the problem of DSAHT and DM-MAC-NF capacity. We summarize their problem statement and results as follows. 

They first reformulate the above problem \textbf{(P1)} into an equivalent optimization problem.
Using the ``common agent'' methodology for decentralized dynamic team problems~\cite{NaMaTe13}, they first decompose the encoding process $x^i_t=f^i_t(m^i,y_{1:t-1})$ into an equivalent two-stage process.
In the first stage, based on the {common information $y_{1:t-1}$}, the mappings (or ``partial encoding functions'') $e^i_t$, $i=1,2$ are
generated as $e^i_t=\phi^i_t[y_{1:t-1}]$\footnote{We use square brackets to denote functions with range being function sets, i.e., we use notation $e^i_t=\phi^i_t[y_{1:t-1}]$ because $e^i_t$ is itself a function.} (or collectively, $e_t=(e^1_t,e^2_t)=\phi_t[y_{1:t-1}]$) where $e^i_t : \mathcal{M}^i \rightarrow \cX^i$. In the second stage, each of these mappings are evaluated at the {private information of each agent}, producing $x^i_t=e^i_t(m^i)$.
In other words, for $i=1,2$, let $\cE^i$ be the collection of  all (deterministic) encoding functions $e^i: \mathcal{M}^i \rightarrow \mathcal{X}^i$.
In the first stage, the common information given by $y_{1:t-1}$ is transformed using mappings $\phi^i_t: \cY^{t-1}  \rightarrow \cE^i$ to produce a pair of encoding functions $e_t=(e^1_t,e^2_t)$. In the second stage these functions are evaluated at the private messages $m^i$ producing $x^i_t=e^i_t(m^i)=\phi^i_t[y_{1:t-1}](m^i)$. 

For the DSAHT problem, authors first define a common belief $\pi_t(m) = P^{\phi}(m|y_{1:t})$ and show that there exists an update function $F$ of $\pi_t$, independent of $\phi$, such that
\eq{
\label{eq:F_update}
\pi_{t} = F(\pi_{t-1},e_{t},y_t)
}
where more explicitly, $F$ is given by
\begin{subequations}\label{eq:pi_update}
\begin{align}
\pi_t&(m^1,m^2) =\frac{Q(y_t|e^1_t(m^1),e^2_t(m^2))\pi_{t-1}(m^1,m^2)}
        {\sum\limits_{m^1,m^2} Q(y_t|e^1_t(m^1),e^2_t(m^2))\pi_{t-1}(m^1,m^2)}
\end{align}
\end{subequations}
Then DSAHT problem can be solved by the following DP,
\eq{
V_{T+1}(\pi_T)&= \E^{\phi}[ 1- \max_{m^1,m^2} \pi_T(m^1,m^2)],\\
V_t(\pi_{t-1}) &= \min_{e_t}\E [  V_{t+1}(F(\pi_{t-1},e_t,Y_t)) | \pi_{t-1} , e_t ] \\
               &= \min_{e_t} \sum_{y_t,m^1,m^2} Q(y_t|e_t^1(m^1),e_t^2(m^2)) \pi_{t-1}(m^1,m^2) \nonumber \\
               &\qquad\qquad \qquad\qquad V_{t+1}(F(\pi_{t-1},e_t,y_t)).
}

Thereafter, regarding the capacity achieving problem, the authors in~\cite{AnPr20} note that the problem of evaluating capacity involves evaluating $C_n(\underline{\lambda})$ for every $\underline{\lambda}$ and is therefore at least as hard as the problem of evaluating the quantity $C_n(\underline{\lambda})$. 
The optimization problem involved in evaluating $C_n(\underline{\lambda})$ can be thought of as a decentralized optimization problem involving two agents: the first is choosing the distribution $q^1_t$ on $x^1_t$
after observing the common information $y_{1:t-1}$ and her private information $x^1_{1:t-1}$, while the second
is choosing the distribution $q^2_t$ on $x^2_t$
after observing the common information $y_{1:t-1}$ and her private information $x^2_{1:t-1}$. They further
show that the capacity expression in~\eqref{eq:region_to_cost_fb} can be expressed as follows.
They first define a private belief $\hat{\pi}^i_t$ as the marginal belief that user $i$ maintains on her own message $m^i$, given her information $(x^i_{1:t},y_{1:t})$ till time $t$, i.e.
\begin{equation}
\hat{\pi}^i_t(m^i) \triangleq \P^{g}(M^i=m^i|x^i_{1:t},y_{1:t}), \qquad i=1,2.
\end{equation}
Note that this belief does not depend on \emph{all} the information available to each transmitter, and specifically does not depend on their own messages $m^i$. Furthermore, they show that there exists functions $\hat{F}^i$ independent of the policies of the transmitters $g$ such that 
\begin{equation}
\hat{\pi}_t^i =
\hat{F}^i(\hat{\pi}^i_{t-1},e^i_t,x^i_t),
\end{equation}
where more explicitly $\hat{F}^i$ is given by,
\begin{align}
\hat{\pi}^i_t(m^i) = \frac{1_{e^i_t(m^i)}(x^i_t)\hat{\pi}^i_{t-1}(m^i)}{\sum_{\tilde{m}^i} 1_{e^i_t(\tilde{m}^i)}(x^i_t)\hat{\pi}^i_{t-1}(\tilde{m}^i)},
\end{align} 
Although the authors do not specify, the repeated application of the above lemma implies that \eq{
\label{eq:pihat_def}
\hat{\pi}^i_t(m^i) = \P^{g}(m^i=m^i|x^i_{1:t}) = \P(m^i=m^i|e^i_{1:t},x^i_{1:t}),
}
i.e. the belief $\hat{\pi}_t^i$ doesn't depend on $y_{1:t}$. Based on this, they derive simplified expressions for the mutual information quantities that are involved in the $I_n(\underline{\lambda})$ in~\eqref{eq:region_to_cost_fb}. Specifically, they derive simplified expressions for the quantities $I(X^1_t;Y_t|X^2_{1:t},Y_{1:t-1})$, $I(X^2_t;Y_t|X^1_{1:t},Y_{1:t-1})$, and $I(X^1_t,X^2_t;Y_t|Y_{1:t-1})$, or equivalently, for the quantities $H(Y_t|X^2_{1:t},Y_{1:t-1})$, $H(Y_t|X^1_{1:t},Y_{1:t-1})$,
$H(Y_t|Y_{1:t-1})$ and $H(Y_t|X^1_t,X^2_t)$. Their results are summarized in the following theorem.
\begin{fact}[Anastasopoulos and Pradhan, 2020]\label{th:capacity}
The mutual information quantities that are involved in the expression for $I_n(\underline{\lambda})$ in ~\eqref{eq:region_to_cost_fb} can be evaluated as expectations of time invariant quantities depended only on $\Pi_{t-1}$, $\hat{\Pi}^i_{t-1}$ and $E_t$. Specifically, for each $t=1,\ldots,n$,
\begin{subequations}
\begin{align}
\label{eq:th2a}
I(X^1_t;Y_t|X^2_{1:t},Y_{1:t-1}) &= \E^{\theta}[ i_1(\hat{\Pi}^2_{t-1},\Pi_{t-1},E_t) ] \\
I(X^2_t;Y_t|X^1_{1:t},Y_{1:t-1}) &= \E^{\theta}[ i_2(\hat{\Pi}^1_{t-1},\Pi_{t-1},E_t) ] \\
I(X^1_t,X^2_t;Y_t|Y_{1:t-1}) &= \E^{\theta}[ i_3(\Pi_{t-1},E_t) ],
\end{align}
\end{subequations}
where the functions $i_1$, $i_2$, $i_3$ are specified in the proof of the theorem and
expectations are taken wrt the joint distribution
\begin{subequations}
\begin{align}
&\P^{\theta}(\pi_{0:n-1},\hat{\pi}_{0:n-1},e_{1:n}) \nonumber \\
&= \prod_{t=0}^{n-1} \P^{\theta}(\pi_t,\hat{\pi}_t,e_{t+1}|\pi_{0:t-1},\hat{\pi}_{0:t-1},e_{1:t}) \\
&=
\prod_{t=0}^{n-1} 1_{\theta_{t+1}[\pi_t]}(e_{t+1}) \sum_{y_t,x^1_t,x^2_t}
  Q(y_t|x^1_t,x^2_t) 1_{F(\pi_{t-1},e_t,y_t)}(\pi_t) \nonumber \\
  &\quad  1_{\hat{F}^1(\hat{\pi}^1_{t-1},e^1_t,x^1_t)}(\hat{\pi}^1_t) 1_{\hat{F}^2(\hat{\pi}^2_{t-1},e^2_t,x^2_t)}(\hat{\pi}^2_t) \nonumber \\
  &\quad \sum_{m^1,m^2}1_{e^1_t(m^1)}(x^1_t)1_{e^2_t(m^2)}(x^2_t) \hat{\pi}^1_{t-1}(m^1)\hat{\pi}^2_{t-1}(m^2).
\end{align}
\end{subequations}
\end{fact}

Equivalently, for a fixed $\underline{\lambda}\in\Real^3_+$, Fact~\ref{th:capacity} shows that the expression $I_n(\underline{\lambda})$ in~\eqref{eq:region_to_cost_fb} involved in evaluating the channel capacity
can be expressed as
\begin{equation}
\label{eq:idef}
I_n(\underline{\lambda})=\frac{1}{n} \sum_{t=1}^n \E^{\theta}[i(\Pi_{t-1},\hat{\Pi}_{t-1},E_t;\underline{\lambda})].
\end{equation}
Furthermore, the unstructured optimization problem for finding $C_n(\underline{\lambda})$ in~\eqref{eq:region_to_cost_fb}
can now be restated as
\begin{equation}
\label{eq:Cn}
C_n(\underline{\lambda}) = \sup_{\theta} \frac{1}{n} \sum_{t=1}^n \E^{\theta}[i(\Pi_{t-1},\hat{\Pi}_{t-1},E_t;\underline{\lambda})].
\end{equation}
The authors argue that the above expression hints at thinking of the quantity $C_n(\underline{\lambda})$ as the average reward received
from a dynamical system with a process $(\hat{\Pi}_{t-1},\Pi_{t-1})$ partially controlled by the encoding functions $E_t=\theta_t[\Pi_{t-1}]$, and optimized over all such policies (Note however that $(\hat{\Pi}_{t-1},\Pi_{t-1})$ is not observed by any single agent, and thus there does not exist a DP to find capacity).
%
The authors also argue that there exists a capacity achieving distribution that is Markovian in the sense of DSAHT problem, based on the following argument. Consider a capacity achieving sequence of transmission schemes indexed by the code length (horizon) n with message size $M_n$ and encoding/decoding functions $e_n, d_n$. Clearly we have a sequence of systems indexed by $n$ such that they achieve the capacity i.e, their rate $R_n := (log_2 M_n) /n \rightarrow C$, and their corresponding error probability $P_e(n) \rightarrow 0$. Now for each element of this sequence with a given $n, M_n$ one can design an optimal scheme for the DSAHT problem. The optimal scheme for a system $n,M_n$ does not change its rate but improves its error probability.

\section{Dynamic program for DM-MAC-NF capacity}
\label{sec:DP}
In this section, we show that there indeed exists a state and consequently a dynamic program for the capacity achieving problem. 
For any policy $\phi$ of the transmitters, we define a new common belief
\eq{\tilde{\pi}_t(m,\hat{\pi}_t) := P^{\phi}(m,\hat{\pi}_{t}|y_{1:t}).
}
Note that $\pi_{t}$ which is a common  belief on $m$ (as defined in the previous section) can be derived from $\tilde{\pi}_t$ as a marginal. As discussed in the previous section, it was shown in~\cite{AnPr20} that there exists capacity achieving policy of the transmitters that is also optimal for DSAHT and is thus of the kind $x_t^i= \theta^i[\pi_{t-1}](m^i)=f^i_t(\pi_{t-1},m^i)$. Thus, we will also restrict ourselves to class of such policies which depend on private information of the player $i$ only through $m^i$. In other words, we \emph{do not} consider policies that depend on player $i$'s complete private information in this set up which is $(\hat{\pi}^i_t,m^i)$, but only on part of it which is $m^i$, as considered in the previous section.
In the following lemma, we show that $\tilde{\pi}_{t-1}$ can be updated using Bayes' rule

\begin{lemma}
There exists functions $\tilde{F}$ independent of the policies of the transmitter $\phi$ such that 
\begin{equation}
\label{eq:Ftilde_update}
\tilde{\pi}_t = \tilde{F}(\tilde{\pi}_{t-1},e_t,y_t)
\end{equation}
\end{lemma}

\begin{IEEEproof}
\eq{
&\tilde{\pi}_t(m,\hat{\pi}_t)\nonumber=P^{\phi}(m,\hat{\pi}_{t}|y_{1:t})\\
&= \frac{\sum_{\hat{\pi}_{t-1}}P^{\phi}(m,\hat{\pi}_{t-1},y_t,\hat{\pi}_{t}|y_{1:t-1})}{\sum_{m,\hat{\pi}_{t-1},\hat{\pi}_{t}}P^{\phi}(m,\hat{\pi}_{t-1},y_t,\hat{\pi}_{t}|y_{1:t-1})}\\
&= \frac{\sum_{\hat{\pi}_{t-1}}\substack{\displaystyle P^{\phi}(m,\hat{\pi}_{t-1}|y_{1:t-1})Q(y_t|e_t^1(m^1),e^2_t(m^2))\\ \displaystyle 1_{\hat{F}(\hat{\pi}^1_{t-1},e^1_t,e_t^1(m^1))}(\hat{\pi}_{t}^1)1_{\hat{F}(\hat{\pi}^2_{t-1},e_t^2,e_t^2(m^2))}(\hat{\pi}_{t}^2)}}{\sum_{m,\hat{\pi}_{t-1},\hat{\pi}_{t}}\substack{\displaystyle P^{\phi}(m,\hat{\pi}_{t-1}|y_{1:t-1})Q(y_t|e_t^1(m^1),e^2_t(m^2))\\ \displaystyle 1_{\hat{F}(\hat{\pi}^1_{t-1},e^1_t,e_t^1(m^1))}(\hat{\pi}_{t}^1)1_{\hat{F}(\hat{\pi}^2_{t-1},e_t^2,e_t^2(m^2))}(\hat{\pi}_{t}^2)}}\\
&= \frac{\sum_{\hat{\pi}_{t-1}}\substack{\displaystyle \tilde{\pi}_{t-1}(m,\hat{\pi}_{t-1})Q(y_t|e_t^1(m^1),e^2_t(m^2))\\\displaystyle 1_{\hat{F}(\hat{\pi}^1_{t-1},e^1_t,e_t^1(m^1))}(\hat{\pi}_{t}^1)1_{\hat{F}(\hat{\pi}^2_{t-1},e_t^2,e_t^2(m^2))}(\hat{\pi}_{t}^2)}}
{\sum_{m,\hat{\pi}_{t-1},\hat{\pi}_{t}}\substack{\displaystyle \tilde{\pi}_{t-1}(m,\hat{\pi}_{t-1})Q(y_t|e_t^1(m^1),e^2_t(m^2))\\
\displaystyle 1_{\hat{F}(\hat{\pi}^1_{t-1},e^1_t,e_t^1(m^1))}(\hat{\pi}_{t}^1)1_{\hat{F}(\hat{\pi}^2_{t-1},e_t^2,e_t^2(m^2))}(\hat{\pi}_{t}^2)}}\\
&= \frac{\pi_{t-1}(m)Q(y_t|e_t^1(m^1),e^2_t(m^2))}
{\sum_{m} \pi_{t-1}(m)Q(y_t|e_t^1(m^1),e^2_t(m^2))}\times\nonumber\\
&\frac{\sum_{\hat{\pi}_{t-1}}\substack{\displaystyle \tilde{\pi}_{t-1}(\hat{\pi}_{t-1}|m) 1_{\hat{F}(\hat{\pi}^1_{t-1},e^1_t,e_t^1(m^1))}(\hat{\pi}_{t}^1)\\\displaystyle 1_{\hat{F}(\hat{\pi}^2_{t-1},e_t^2,e_t^2(m^2))}(\hat{\pi}_{t}^2)}}
{\sum_{\hat{\pi}_{t-1},\hat{\pi}_{t}}\substack{\displaystyle \tilde{\pi}_{t-1}(\hat{\pi}_{t-1}|m) 1_{\hat{F}(\hat{\pi}^1_{t-1},e^1_t,e_t^1(m^1))}(\hat{\pi}_{t}^1)\\\displaystyle1_{\hat{F}(\hat{\pi}^2_{t-1},e_t^2,e_t^2(m^2))}(\hat{\pi}_{t}^2)}} \label{eq:L1a}
}
\end{IEEEproof}


The next step in the development is to derive simplified expressions for the mutual information quantities that are involved in the $I_n(\underline{\lambda})$ in~\eqref{eq:region_to_cost_fb}. Specifically, we will derive simplified expressions for the quantities $I(X^1_t;Y_t|X^2_{1:t},Y_{1:t-1})$, $I(X^2_t;Y_t|X^1_{1:t},Y_{1:t-1})$, and $I(X^1_t,X^2_t;Y_t|Y_{1:t-1})$, or equivalently, for the quantities $H(Y_t|X^2_{1:t},Y_{1:t-1})$, $H(Y_t|X^1_{1:t},Y_{1:t-1})$,
$H(Y_t|Y_{1:t-1})$ and $H(Y_t|X^1_t,X^2_t)$. Our results are summarized in the following theorem.
\begin{lemma}\label{th:capacity2}
The mutual information quantities that are involved in the expression for $I_n(\underline{\lambda})$ in~\eqref{eq:region_to_cost_fb} can be evaluated as expectations of time invariant quantities depended only on $\tilde{\Pi}_{t-1}$, and $E_t$. Specifically, for each $t=1,\ldots,n$ we have
\begin{subequations}
\label{eq:th2a}
\begin{align}
I(X^1_t;Y_t|X^2_{1:t},Y_{1:t-1}) &= \E^{\theta}[ \tilde{i}_1(\tilde{\Pi}_{t-1},E_t) ] \\
I(X^2_t;Y_t|X^1_{1:t},Y_{1:t-1}) &= \E^{\theta}[ \tilde{i}_2(\tilde{\Pi}_{t-1},E_t) ] \\
I(X^1_t,X^2_t;Y_t|Y_{1:t-1}) &= \E^{\theta}[ \tilde{i}_3(\tilde{\Pi}_{t-1},E_t) ],
\end{align}
\end{subequations}
where the functions $i_1$, $i_2$, $i_3$ are specified in the proof of the theorem and
expectations are taken wrt the joint distribution
\begin{subequations}
\begin{align}
&\P^{\theta}(\tilde{\pi}_{0:n-1},e_{1:n}) = \prod_{t=0}^{n-1} \P^{\theta}(\tilde{\pi}_t,e_{t+1}|\tilde{\pi}_{0:t-1},e_{1:t}) \\
&=
\prod_{t=0}^{n-1} 1_{\theta_{t+1}[\tilde{\pi}_t]}(e_{t+1})\times \nonumber\\
&\sum_{y_t,m^1,m^2}\pi_t(m^1,m^2)
  Q(y_t|e^1_t(m^1),e^2_t(m^2))1_{\tilde{F}(\tilde{\pi}_{t-1},e_t,y_t)}(\tilde{\pi}_t) 
  \end{align}
\end{subequations}
\end{lemma}

\begin{IEEEproof}
Let $i_1(\hat{\Pi}^2_{t-1},\Pi_{t-1},E_t), i_2(\hat{\Pi}^1_{t-1},\Pi_{t-1},E_t)$, $i_3(\Pi_{t-1},E_t)$ be as defined in Theorem~1 in~\cite{AnPr20}. Define 
\seq{
\eq{
\tilde{i}_1(\tilde{\pi}_{t-1},e)&= \sum_{\hat{\pi}_{t-1}}i_1(\hat{\pi}_{t-1}^2,\pi_{t-1},e)\tilde{\pi}_{t-1}(\hat{\pi}_{t-1}^2)\\
\tilde{i}_2(\tilde{\pi}_{t-1},e)&= \sum_{\hat{\pi}_{t-1}}i_2(\hat{\pi}_{t-1}^1,\pi_{t-1},e)\tilde{\pi}_{t-1}(\hat{\pi}_{t-1}^1)\\
\tilde{i}_3(\tilde{\pi}_{t-1},e)&= i_3(\pi_{t-1},e),
}
}
and $\tilde{i}(\tilde{\pi}_{t-1},e_t)$ is appropriately defined through the above quantities in a similar way $i(\pi_{t-1},\hat{\pi}_{t-1},e_t)$ in~\eqref{eq:idef} is defined through quantities in~\eqref{eq:th2a}.
\optv{arxiv}{

Consequently, the mutual information quantities at time $t$ become
\begin{subequations}
\begin{align}
I&(X^1_t;Y_t|X^2_{1:t},Y_{1:t-1})
 = \E^{\theta}[ \tilde{i}_1(\tilde{\Pi}_{t-1},E_t) ] \\
I&(X^2_t;Y_t|X^1_{1:t},Y_{1:t-1})
  = \E^{\theta}[ \tilde{i}_2(\tilde{\Pi}_{t-1},E_t) ] \\
I&(X^1_t,X^2_t;Y_t|Y_{1:t-1})
 = \E^{\theta}[ \tilde{i}_3(\tilde{\Pi}_{t-1},E_t) ]
\end{align}
\end{subequations}

}
\end{IEEEproof}

Based on Lemma~2 we can now solve for the capacity of DM-MAC-NF as a dynamic program for the common agent as follows. 
\begin{theorem}
The dynamic optimization problem in~\eqref{eq:Cn} can be solved using the following dynamic program

\begin{subequations}
\label{eq:DP2}
\begin{align}
&J+V(\tilde{\pi})  = \max_{e}  \E[\tilde{i}(\tilde{\pi},e;\underline{\lambda}) +  V(\tilde{F}(\tilde{\pi},e,Y)) | \tilde{\pi} , e ] \\
          &= \max_{e}  \tilde{i}(\tilde{\pi},e;\underline{\lambda}) +  \sum_{y,m^1,m^2}{\pi}(m^1,m^2)\times\nonumber\\
          & Q(y|e^1(m^1),e^2(m^2)) V(\tilde{F}(\tilde{\pi},e,y)).
\end{align}
\end{subequations}
\end{theorem}

\begin{IEEEproof}
We note that $\{\tilde{\pi}_t,e_t\}_t$ is a controlled Markov process for this problem, since the instantaneous cost can be written as a function $\tilde{\pi}_t,e_t$, and $P(\tilde{\pi}_{t+1}|\tilde{\pi}_{1:t},e_{1:t})= P(\tilde{\pi}_{t+1}|\tilde{\pi}_{t},e_{t})$, as implied by~\eqref{eq:Ftilde_update}. Thus the result is implied by Markov decision process theory~\cite{KuVa86}.
\end{IEEEproof}
The above result implies that there exist optimal Markovian policy of the transmitter $i$ that is a function of the state $\tilde{\pi}_t$ and $m^i$.

\subsection{Simplification}
We recall that it was shown in~\cite{AnPr20} that there exists capacity achieving policies of the players that depend only on $\pi_t,m^i$, (and not on $\tilde{\pi}_t, m^i)$. However, Theorem~1 above shows that there exists an optimal strategy through the DP that depends on $(\tilde{\pi}_t, m^i)$. Here, we show that there exists a one to one function between $\pi_t$ and $\tilde{\pi}_t$, and thus one can restrict oneself to a smaller space of $\pi_t$ and yet derive the belief $\tilde{\pi}_t$. To show the equivalency, we note that 
\eq{
\tilde{\pi}_t(m,\hat{\pi}_t) &:= P^{\phi}(m,\hat{\pi}_{t}|y_{1:t})\\
&:= P^{\phi}(m|y_{1:t})P^{\phi}(\hat{\pi}_{t}|m,y_{1:t})\\
&= \pi_t(m)\tilde{\pi}_t(\hat{\pi}_t|m)
}
Now note that $\hat{\pi}_t^i$ is a function of $(e_{1:t}^i,x_{1:t}^i) = (e_{1:t}^i,e_{1:t}^i(m^i))$ as shown in~\eqref{eq:pihat_def}. Thus knowing $m,\phi,y_{1:t}$ and equivalently $m,e_{1:t}$, $\hat{\pi}_t$ is perfectly observed i.e. $\tilde{\pi}_t(\hat{\pi}_t|m) = \delta(\hat{\pi}_t(\cdot) = P(\cdot|e_{1:t},e_{1:t}(m)))$. This can also be seen in~\eqref{eq:L1a} where through induction if $\tilde{\pi}_{t-1}(\hat{\pi}_{t-1}|m)$ is a delta function then so is $\tilde{\pi}_{t}(\hat{\pi}_{t}|m)$.
%
%
Thus one can reduce the state space in DP in~\eqref{eq:DP2} from $\tilde{\pi}_t$ to $\pi_t$.

\begin{theorem}
The dynamic optimization problem in~\eqref{eq:Cn} can be solved using the following dynamic program
\begin{align}
&J+V(\pi)  = \max_{e}  \E[\tilde{i}(\tilde{\pi},e;\underline{\lambda}) +  V(F(\pi,e,Y)) | \pi , e ] 
\end{align}
where as argued before $\tilde{\pi}_t$ can be derived from $\pi_t$.

\end{theorem}
\begin{IEEEproof}
The proof is implied from Theorem~1 and the above discussion.
\end{IEEEproof}

\section{Conclusion and future work}
\label{sec:Conl}
In this paper, we considered the problem of finding capacity of discrete memoryless multiple access channel with noiseless feedback. In~\cite{AnPr20}, authors made connections with finding capacity of this channel with the problem of minimizing the terminal probability of error, and show that achieving capacity is a dynamic optimization problem. We build upon their work and show that there exists a state and a dynamic program to find the capacity of this channel. Thus we show that there is a single letter characterization of the capacity expression whose alphabet is a belief state.

Future work involves deriving the already known results in this domain using this new framework such as~\cite{Oz84} for Gaussian MAC with feedback, for which the capacity expression is known, or for Cover-Leung channels~\cite{CoLe81} again for which the capacity is known, or sum capacity of N-player Gaussian channel~\cite{SuGaKr19}, and building upon that.

\bibliographystyle{IEEEtran}

\begin{thebibliography}{10}
\providecommand{\url}[1]{#1}
\csname url@samestyle\endcsname
\providecommand{\newblock}{\relax}
\providecommand{\bibinfo}[2]{#2}
\providecommand{\BIBentrySTDinterwordspacing}{\spaceskip=0pt\relax}
\providecommand{\BIBentryALTinterwordstretchfactor}{4}
\providecommand{\BIBentryALTinterwordspacing}{\spaceskip=\fontdimen2\font plus
\BIBentryALTinterwordstretchfactor\fontdimen3\font minus
  \fontdimen4\font\relax}
\providecommand{\BIBforeignlanguage}[2]{{%
\expandafter\ifx\csname l@#1\endcsname\relax
\typeout{** WARNING: IEEEtran.bst: No hyphenation pattern has been}%
\typeout{** loaded for the language `#1'. Using the pattern for}%
\typeout{** the default language instead.}%
\else
\language=\csname l@#1\endcsname
\fi
#2}}
\providecommand{\BIBdecl}{\relax}
\BIBdecl

\bibitem{Kr98}
G.~Kramer, ``{Directed information for channels with feedback},'' \emph{Ph.D.
  dissertation, ETH Series in Information Processing. Konstanz, Switzerland:
  Hartung-Gorre Verlag,}, 1998.

\bibitem{AnPr20}
A.~Anastasopoulos and S.~Pradhan, ``{Decentralized sequential active hypothesis
  testing and the MAC feedback capacity},'' \emph{IEEE International Symposium
  on Information Theory - Proceedings}, vol. 2020-June, pp. 2085--2090, jun
  2020.

\bibitem{GaKi11}
Y.-H. {El Gamal, Abbas and Kim}, \emph{{Network information theory}}, 2011.

\bibitem{GaWo75}
N.~T. Gaarder and J.~K. Wolf, ``{The Capacity Region of a Multiple-Access
  Discrete Memoryless Channel Can Increase with Feedback},'' \emph{IEEE
  Transactions on Information Theory}, vol.~21, no.~1, pp. 100--102, 1975.

\bibitem{CoLe81}
T.~M. Cover and C.~S. Leung, ``{An Achievable Rate Region for the
  Multiple-Access Channel with Feedback},'' \emph{IEEE Transactions on
  Information Theory}, vol.~27, no.~3, pp. 292--298, 1981.

\bibitem{Oz84}
L.~H. Ozarow, ``{The Capacity of the White Gaussian Multiple Access Channel
  with Feedback},'' \emph{IEEE Transactions on Information Theory}, vol.~30,
  no.~4, pp. 623--629, 1984.

\bibitem{NaMaTe13}
A.~Nayyar, A.~Mahajan, and D.~Teneketzis, ``Decentralized stochastic control
  with partial history sharing: A common information approach,''
  \emph{Automatic Control, IEEE Transactions on}, vol.~58, no.~7, pp.
  1644--1658, 2013.

\bibitem{Sh48}
C.~E. Shannon, ``A mathematical theory of communication,'' \emph{Bell system
  technical journal}, vol.~27, no.~3, pp. 379--423, 1948.

\bibitem{Sa78}
M.~Salehi, ``{Cardinality bounds on auxiliary variables in multiple-user theory
  via the method of Ahlswede and K{\{}$\backslash$"o{\}}rner},'' \emph{Dept.
  Statistics, Stanford Univ., Stanford, CA, Tech. Rep}, vol.~33, 1978.

\bibitem{KuVa86}
P.~Kumar and P.~Varaiya, ``Stochastic systems,'' 1986.

\bibitem{SuGaKr19}
E.~Sula, M.~Gastpar, and G.~Kramer, ``{Sum-Rate Capacity for Symmetric Gaussian
  Multiple Access Channels with Feedback},'' \emph{IEEE Transactions on
  Information Theory}, vol.~66, no.~5, pp. 2860--2871, may 2020.

\end{thebibliography}

\end{document}